\documentclass[aps,superscriptaddress,amsmath,showpacs,twocolumn,pra]{revtex4-2}
\usepackage[T1]{fontenc}
\usepackage[latin9]{inputenc}
\usepackage{amssymb}
\usepackage{graphicx}
\usepackage{subfigure}
\usepackage{epsfig}
\usepackage{txfonts}
\usepackage{amsfonts}
\usepackage{dcolumn}
\usepackage{esint}
\usepackage{mathrsfs}
\usepackage{enumerate}
\usepackage{color}
\begin{document}

\title{
Wide-range quantum enhanced rotation sensing with 1+2 dimensional dynamical decoupling}
\author{X. N. Feng}
\affiliation{Information Quantum Technology Laboratory, International Cooperation Research Center of China Communication and Sensor Networks for Modern Transportation, School of Information Science and Technology, Southwest Jiaotong University, Chengdu 610031, China}
\author{L. F. Wei}
\email{Corresponding author. lfwei@swjtu.edu.cn}
\affiliation{Information Quantum Technology Laboratory, International Cooperation Research Center of China Communication and Sensor Networks for Modern Transportation, School of Information Science and Technology, Southwest Jiaotong University, Chengdu 610031, China}

\begin{abstract}
We propose a motional dynamical decoupling technique by utilizing a sequence of $\pi$-phase shifts, instead of the conventional $\pi$-pulses for spin flipping, to implement the quantum enhanced rotation sensing with a 1+2 dimensional hybrid atomic Sagnic interferometor. By fully disentangling the spin from the two-dimensional vibrational modes of the particle under rotation,  the spin coherence time and thus the  phase accumulation can be significantly increased. Consequently, both the achievable sensitivity and dynamic range of the rotation sensing can be significantly enhanced and extended simultaneously, compared to the previous schemes where the spin and motions of the particle were not completely decoupled. The experimental feasibility for the unambiguous estimation of the rotation parameters is also discussed. Hopefully, this technique holds promise for overcoming certain challenges existing in the usual matter-wave Sagnac interferometers with trapped particles,  particularly for the practical inertial navigation that demands both high sensitivity and large dynamic range.

\end{abstract}
\maketitle

{\it Introduction.---}
High sensitive and wide dynamic range of rotation sensings are crucial for inertial navigation~\cite{Alzar2019,Bongs2019}, geodesy~\cite{Bongs2019,Citation2014}, and testing the effects of relativity~\cite{Will2006}, etc.. Among the various available sensing approaches, atomic Sagnac interferometer (ASI) possesses certain specific advantages, due to their high sensitivities and stabilities, and relatively easy experimental demonstrations~\cite{Lenef1997}. Particularly, the
recent advances with the atoms confined in the guiding potentials or traps~\cite{Saijun2007,Stevenson2015,YanH2012,Navez2016,Campbell2017,Helm2018,Carlos2019,Beydler2024,Dash2023,Moan2020,Feng2022,Feng2023,Wang2023}, rather than freely falling through space, suggest that
the constraints of the substantial size and complexity of the experimental setups can be further alleviated. Despite the remarkable potential of ASIs utilizing trapped or guided atoms for high-sensitivity rotation sensing, several challenges remain. For instance, increasing the enclosed area by employing the larger macro-superposition states with greater spatial separation might introduce certain stronger background noises~\cite{Campbell2017,Wang2023}, which would limit the coherent time of the atomic system and thus its achievable measurement sensitivity. Moreover, the enhancement of sensitivity is often at the expense of the decrease of the dynamic range of the rotation parameter. Therefore, developing an ASI technique that achieves both high sensitivity and a wide dynamic range remains a significant challenge~\cite{Beydler2024,Moan2020,Dash2023}, particularly for the practical inertial navigation applications that require both features.

To address such an issue for basically leveraging the potential advantages of the ASIs, the dynamical decoupling (DD) method or the inverse evolution technique~(see, e.g.,~\cite{Saijun2007,Japha2007,Schubert2021}),
might play an important role, as they can effectively suppress the unwanted coupling noises.
For instance, in a trapped-condensate interferometer, the coherence time can be enlarged by using two $\pi/2$ Bragg pulses to suppress the dephasing~\cite{Munekazu2007}. More recently, a motional dynamical decoupling (MDD) technique employing a sequence of $\pi$-pulses has been introduced for sensitive weak force sensing, by protecting the desired coherence from the environmental perturbation~\cite{Pedernales2020}.
However, these methods are typically limited to the 1+1 dimension system-i.e., a 1/2-spin coupling to a one-dimensional harmonic oscillator, such as the atoms confined in a ring dipole potential~\cite{Bell2016}, magnetic trap~\cite{Qi2017}, and the time-averaged adiabatic potentials~\cite{Navez2016}, etc.. In the context of rotation sensing, these models often oversimplify the system by assuming infinite radial confinement.
In practice, the non-negligible radial force, which depends on both the rotational-speed and spin-state~\cite{Stevenson2015}, would inevitably introduce certain unwanted noises. Thus, a practical trapped-particle-based Sagnac interferometer (TPSI) should involve a two-dimensional~\cite{Moan2020,Feng2022,Wang2023} or even three-dimensional harmonic oscillator~\cite{Shinjo2021}.
Unfortunately, the above MDD technique, though applied successfully to the usual 1+1 dimensional atomic interferometors for one-dimensional rotation and force sensing, can not be directly applied to the present 1+2 dimensional TPSIs for the sensitive rotation sensing, due to the rotation-induced coupling between different vibrational degrees of freedoms of the particle~\cite{Campbell2017}.

Inspired by the success of the previous MDD  technique, which can be implemented simply by using the $\pi$-pulse sequences for spin flipping in 1+1 dimensional atomic interferomters, here we propose an $\pi$-phase shift based MDD technique (called as PMDD for simplicity).
Instead of using the $\pi$-pulses, the PMDD technique employs a series of $\pi$-phase shift operations to achieve complete decoupling between the spin and the two-dimensional vibrational modes of a trapped atom in the current 1+2 dimensional TPSI.
With the proposed PMDD method, we show that the phase accumulation time of the atomic spin can be significantly increased for arbitrary rotation speeds, thereby enabling high-precision rotation sensing over a much wider dynamic range. Hopefully, the proposed technique  might be helpful for the implementation of  the practical inertial navigation, where both high sensitivity and a broad dynamic range are essential.

\begin{figure}[htbp]
    \centering
    \includegraphics[width=0.95\linewidth]{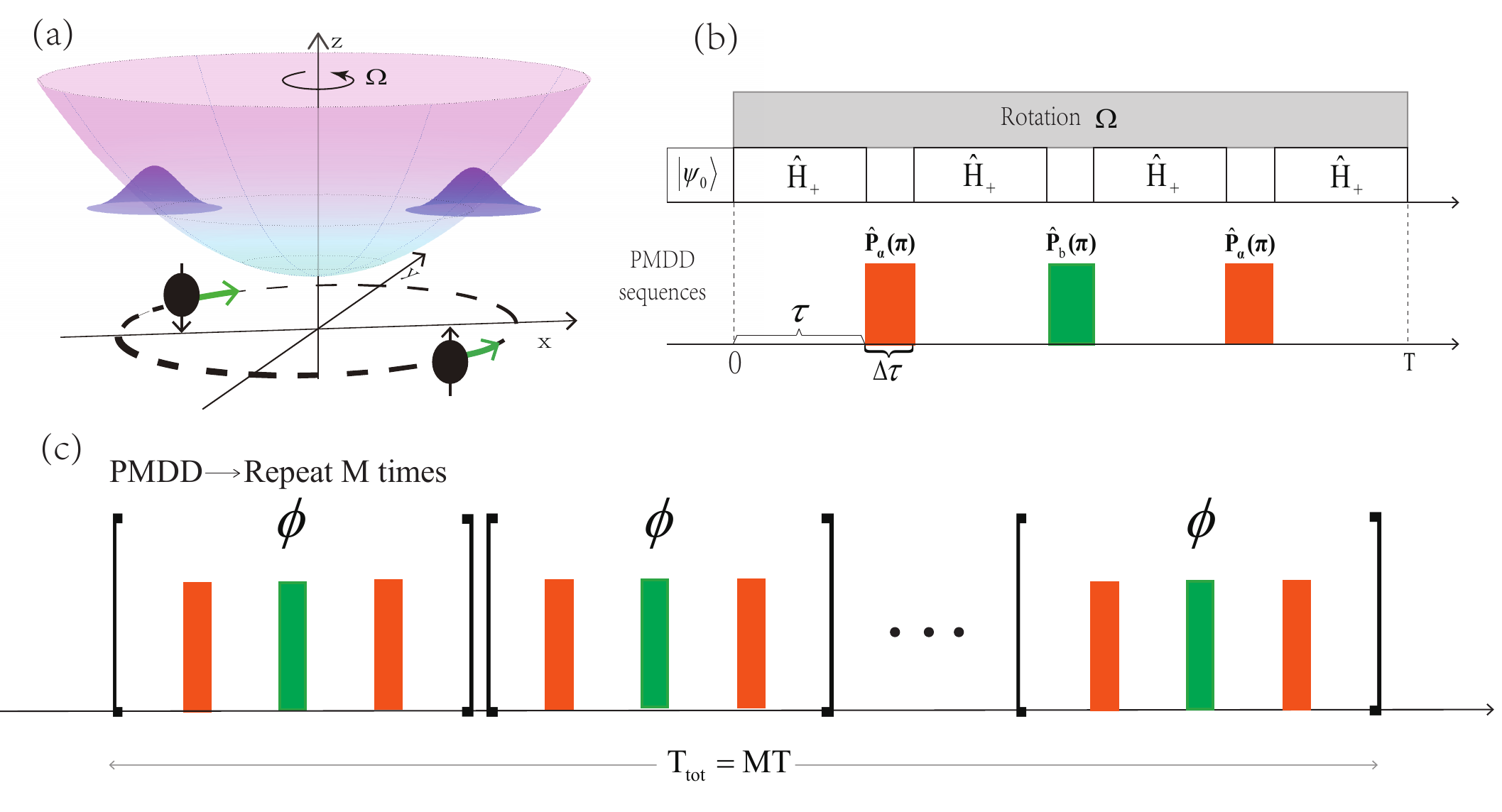}
    \caption{A 1+2 dimensional atomic Sagnac interferometer and the proposed PMDD technique for sensitive rotation sensings. (a) A single particle is trapped in a two-dimensional harmonic potential and rotates with an external rotation speed $\Omega$, which is measured by the projective detection of the atomic spin, i.e., the internal state of the trapped particle.  (b) The operational unit of the proposed PMDD technique, which is consisted of a sequence of $\pi$-phase shifts $\hat{P}_k(\pi)=e^{i\pi\hat{k}^\dagger\hat{k}}$ on mode-k with k=a,b, instead of the conventional $\pi$-pulses, implemented between two displacement operations $\hat{H}_+$ of Eq.~(\ref{H13}) to completely disentangle the spin from the vibration degree. (c) Repeatedly using the PMDD technique to implement the  desirably continuous accumulation of the relative phase of the atomic spin state.}
    \label{f1}
\end{figure}
{\it Model and Method.---}
Let us consider a generic 1+2 dimensional atomic Sagnac interferometer, shown schematically in Fig.~\ref{f1}. This system consists of a trapped particle with a 1/2-spin coupled to a two-dimensional harmonic oscillator, as described in works such as~\cite{Stevenson2015,Feng2022,Moan2020,Wang2023}. The particle's spin-dependent vibrations in a two-dimensional plane, along the reverse directions, create the two paths of the Sagnac interferometer. Physically, such an atomic Sagnac interferometer configuration can be realized by various trapped particles, typically such as the atom~\cite{Stevenson2015}, ion~\cite{Campbell2017}, electron~\cite{Carney2021}, and also a Bose-Einstein condense~\cite{Helm2018} etc.. Here, the vibration of the trapped particle along the z direction is assumed to be strongly confined for the simplicity. The two-dimensional oscillator vibrating in the $x-y$ plane is served as the probe for the sensing of rotating angular speed $\Omega$, while the $1/2-$spin acts as an ancillary to accumulate the phase and also as the detector for reading out the relevant rotation information.

The generic dynamics for such an atomic Sagnac interferometer can be modeled by the following Hamiltonian~\cite{Campbell2017,Moan2020} ($\hbar=1$)
\begin{equation}\label{H0}
  {\hat{H}_s = \hat{H}_0+\hat{H}_\Omega+\hat{H}_c},
\end{equation}
where $\hat{H}_0=\omega_x\hat{a}^\dagger\hat{a}+\omega_y\hat{b}^\dagger\hat{b}$ describes the two-dimensional harmonic oscillator of atom with the frequency $\omega_{x/y}$ and the annihilation (creation) operators $\hat{a}$ ($\hat{a}^\dagger$) and $\hat{b}$ ($\hat{b}^\dagger$) along the $x$- and $y$- directions, respectively; $\hat{H}_\Omega=\Omega\hat{J}_y$ describes the driven rotation of the particle with $\Omega$ being the rotation speed  and $J_y=(\hat{a}^\dagger\hat{b}-\hat{a}\hat{b}^\dagger)/2$ the angular momentum operator; $\hat{H}_c=\alpha(t)\hat{x}_a\sigma_z+\beta(t)\hat{p}_b$ is the spin-dependent driving. It generates the displacement operation of the vibrational mode-b with amplitude $\beta(t)$, and also the entanglement between the spin and another vibrational mode-a, with the spin-oscillator coupling strength $\alpha(t)$. Above, $\hat{x}_k=\hat{k}^\dagger+\hat{k}$ and $\hat{p}_k=i(\hat{k}^\dagger-\hat{k})$ denote the position and momentum operators of the mode $k$, with $k=a,b$, respectively.

Primarily, with the present atomic Sagnac interferometer,  the rotation parameter $\Omega$ can be sensitively estimated, in principle, by using the  spin-motion entangled evolution and then the projective measurements performed on the spin state. Indeed,
in the interaction picture, the Hamiltonian of Eq.~(\ref{H0}) can be simplified as
\begin{equation}\label{H13}
  \hat{H}_+ = \Omega\hat{J}_y+\alpha\hat{x}_a\sigma_z+\beta\hat{p}_b,
\end{equation}
with $\alpha(t)=\alpha\cos(\omega t)$ and $\beta(t)=\beta\cos(\omega t)$, under the rotating-wave approximation (RWA).
The relevant  unitary evolution can be obtained as (see Appendix A for details):
\begin{equation}\label{evo1}
\begin{split}
\hat{U}_+(t)
=&e^{i\phi_1\sigma_z}\exp\Big({\alpha t}\left[A_\theta\hat{x}_a+B_\theta\hat{x}_b\right]\sigma_z\Big)\\
\times&\exp\Big({\beta t}\left[A_\theta\hat{p}_b-B_\theta\hat{p}_a\right]\Big)\hat{P}(\theta),
\end{split}
\end{equation}
where $A_\theta={\sin(\theta)}/{\theta}$ ,  $B_\theta={(1-\cos(\theta))}/{\theta}$,  $\hat{P}(\theta)=e^{i\theta J_y}$ with $\theta=\Omega t$. After the above evolution, the system becomes a spin-motion entangled state, i.e., $|\Psi(\Omega)\rangle=\hat{U}_+(t)|\psi_0\rangle$, where $|\psi_0\rangle$ is the initial separable state of the system.
Then, the rotation parameter $\Omega$ can be inferred from the spin population
\begin{equation}\label{population1}
P_\downarrow(\theta)=\frac{1}{2}\left(1+e^{-\Gamma(\theta)}\cos(\phi_1)\right), \phi_1 = -\alpha\beta t^2\theta^{-2}[\sin(\theta)-\theta],
\end{equation}
by directly performing the projective measurements on the spin $|\downarrow\rangle$-state. Here, 
 $\phi_1$
is the accumulated phase, and $\Gamma(\theta)=8(\alpha\sin(\theta/2)/\Omega)^2$.
Obviously, the sensitivity to estimate the rotation parameter $\Omega$ should be very limited, as the measured spin-state population decays by a factor of $\Gamma(\theta)$, except $\sin(\theta/2)\ll1$ (with $\theta\approx k\pi~ (k\in N)$) which imposes a strong constraint on the dynamic range of $\Omega$ for a given duration $t$, or vice versa. Thus, the potential applications involved particularly the relatively large values of $\Omega$ is very limited, typically such as for the practical inertial navigation. Actually, as mentioned above, this is an open problem encountered by most of the TPSIs in 1+2 dimensional scenarios~\cite{Moan2020,Feng2022,Wang2023}, and thus a trade-off exists between the high sensitivity rotation sensing and the wide dynamic range.

To address such a problem, below we propose a new approach, called as the $\pi$-phase-shift-based motional dynamical decoupling (PMDD) technique. This approach involves completely disentangle the spin from the two dimensional harmonic oscillator before the spin projective measurements are performed, thereby enhancing the sensitivity and dynamic range. For this aim, an inverse evolution
\begin{equation}\label{evo2}
\begin{split}
\hat{U}_-(t)
=&\hat{P}(-\theta)\exp\left({-\alpha t}\left[A_\theta\hat{x}_a+B_\theta\hat{x}_b\right]\sigma_z\right)\\
&\times\exp\left({\beta t}\left[A_\theta\hat{p}_b+B_\theta\hat{p}_a\right]\right)e^{i\phi_1}.
\end{split}
\end{equation}
is required, which can be generated by the Hamiltonian (see the appendix A also for details)
\begin{equation}\label{reverse}
  \hat{H}_-  = -\Omega\hat{J}_y-\alpha\hat{x}_a\sigma_z+\beta\hat{p}_b.
\end{equation}
This Hamiltonian can be theoretically realized by rotating the particle in the opposite direction and the spin is flipped simultaneously by using a $\pi$-pulse operation similar to the conventional MDD.
Indeed, after the evolution $\hat{U}_+(t)$ and the inverse evolution $\hat{U}_-(t)$, with the same duration $\tau$, the system is evolved as a spin-motion separated state, i.e.,
\begin{equation}\label{evo3}
\begin{split}
 |\Psi_1(2\tau)\rangle=\hat{U}_-(\tau)\hat{U}_+(\tau)|\psi_0\rangle=e^{{2\beta t}A_\theta\hat{p}_b}e^{2i\phi_1\sigma_z}|\psi_0\rangle,
  \end{split}
\end{equation}
and thus the information of the rotation parameter $\Omega$ can be encoded into the relative phase $\phi_1$ of the spin superposition state for detection.
Noted that, although the spin and motion are decoupled completely, the oscillator is evolved into the coherent states with the increasingly larger displaced amplitudes; i.e., $2\beta tA_\theta$ in Eq.~(\ref{evo3}), in the y-direction vibration. These displacements would increase further with the additional operations $\hat{U}_-(\tau)$ and $\hat{U}_+(\tau)$ during the relevant phase accumulations, and thus may introduce some unexpected noises that would degrade the sensitivity of the rotation sensing.
Fortunately, one can see that, these unwanted displacements can be effectively canceled by simply adding a  $\pi$-phase shift operation $e^{i\pi\hat{N}}$, with $\hat{N}=\hat{a}^\dagger\hat{a}+\hat{b}^\dagger\hat{b}$ being the total photon number operator. Consequently, after a two-step process involving the successive $\hat{U}+(\tau)$ and $\hat{U}-(\tau)$ evolutions, the system evolves as (see Appendix B for details)
\begin{equation}\label{psi2}
|\Psi_2(4\tau)\rangle= \hat{U}_-(\tau)\hat{U}_+(t)e^{i\hat{N}\pi}\hat{U}_-(\tau)\hat{U}_+(t)|\psi_0\rangle=e^{4i\phi_1\sigma_z}e^{i\hat{N}\pi}|\psi_0\rangle,
\end{equation}
which indicates obviously that the two-dimensional harmonic oscillator is now returned into its initial state differing only by a phase. Basically, the present MDD technique is based on a $\pi$-phase shift operation, called consequently as the PMDD for the simplicity, and thus is certainly  different from the previous MDD one used well in the usual 1+1 dimensional atomic interferometor by using the $\pi$-pulses.

Repeating the PMDD evolutions (\ref{psi2}) multiple times within the long coherence time $T=4M\tau$ (with $M$ being the repeated times), the relative phase of the atomic spin state can be continuously accumulated. As a consequence, by the final spin-state projective measurement,  the desired rotation parameter $\Omega$ can be estimated from the probability
\begin{equation}\label{population2}
  P_\downarrow(\Omega,t)=\frac{1}{2}\Big(1+\cos[8M\phi_1(\Omega,t)]\Big),
\end{equation}
for the spin state $|\downarrow\rangle$ projective measurements, with $\phi_1(\Omega,t)=\phi_1$. Compared with Eq.~(\ref{population1}), we see that the measured population demonstrated here does not decay with the increasing phase accumulation, and thus the high measurement fidelity can be achieved for arbitrary $\theta$, at least theoretically.
Indeed, With the increasing phase accumulation,
the present uncertainty of the estimated parameter $\Omega$ reads
\begin{equation}\label{domega}
  \Delta\Omega=\frac{\Delta P_{\downarrow}(\Omega,\tau)}{|\partial P_{\downarrow}(\Omega,\tau)/\partial\Omega|}=\frac{\Omega^3\sqrt{\tau}}{4\sqrt{M}\alpha\beta \big|\theta+\theta\cos(\theta)-2\sin(\theta)\big|}.
\end{equation}
Specifically, it is approximated as $\Delta\Omega\approx 3/2\sqrt{4M\tau}\alpha\beta \tau^2$ for $\theta\leq1 rad$, wherein
$[\theta+\theta\cos(\theta)-2\sin(\theta)]/\theta^3\approx 1/6$.

Compared with the previous results~\cite{Stevenson2015,Campbell2017,Wang2023,Moan2020}, the proposed PMDD technique offers two main advantages in rotation sensing.
Firstly, the PMDD technique enables the implementation of high sensitive rotation sensing for large rotation parameter $\Omega$, i.e., $\theta \in (2l\pi - 1, 2l\pi + 1)$ $l=0,1,2,...$ for a given duration $\tau$. In contrast to earlier approaches~\cite{Campbell2017,Wang2023}, which require the condition $\sin(\theta/2) \ll 1/(2\sqrt{2\bar{N}_c})$ (with $\bar{N}_c$ representing the amplitude of the cat-state) due to incomplete decoupling of the spin and two-dimensional motion, the present dynamic range for the  parameter estimation is independent of the cat-state properties, and thus the effective dynamic range of the rotation parameter can be  significantly extended.
Secondly, with the proposed PMDD technique the additional noises, generated by the phonon excitation during the spin-motion entangled evolution, can be effectively canceled, allowing for an extended phase accumulation time. This enhancement would result in the higher sensitivity of rotation sensing, which is improved by a factor of $1/\sqrt{M}$. More importantly, augmenting $M$ to prolong the phase accumulation time does not compromise the dynamic range.

{\it Physical implementations.---}
We now investigate how to demonstrate the proposed PMDD for the desired rotation sensing with both the high sensitivity and the large parameter dynamic range. Without loss of the generality, let us consider specifically a 1+2 dimensional TPST generated by the trapped $^{171}Yb+$ ion~\cite{Campbell2017}, wherein $(\omega_x, \omega_y , \omega_z) = (10, 10, 1)$ kHz. In presence of an external rotation around the
z-axis, the Hamiltonian of the trapped ion (whose motion being confined effectively in the
$x-y$ plane) can be written as
   $\hat{H}_r = \omega_0\sigma_z/2+\hat{H}_0+\hat{H}_\Omega$,
with $\omega_0$ being the transition frequency between the spin states encoded by the selected two levels of the ion's internal states. To delivers the third term in Eq.~(1), i.e., $\hat{H}_c$, for the generation of spin-motion entanglement, a resonant driving electric field $E(t)=E\cos(\omega t+\phi_y)$ along the y-axis and a spin-dependent force $f(t) = \alpha\cos(\omega t)\sigma_z$ along the x-axis, are required to be applied. Physically, these two drivings  generate a displacement  operation $\beta\cos(\omega t)\hat{p}_b$ on the $y$-direction vibration, and an interaction $\alpha\cos(\omega t)\hat{x}_a\sigma_z$ between the spin with the $x$-direction vibration. Indeed,
these operations had been experimentally realized in the trapped ion platform~\cite{Gilmore2021,Burd2019}, where the amplitudes $\alpha$ and $\beta$ are controllable by simply adjusting the strengths of the applied drivings. Therefore, the Hamiltonian (1) can be realized with the present trapped ion system.

Given the spin-motion entangled operation $\hat{U}_+(t)$ can be directly realized by the Hamiltonian (\ref{H13}), below we focus on how to realize the spin-motion disentangled operation $\hat{U}_-(t)$ and
the evolution $e^{i\pi\hat{k}^\dagger\hat{k}}$ ($ \hat{k}=a,b$), which is crucial to cancel the unwanted phonon excitation.
Theoretically, the evolution $e^{i\pi\hat{k}^\dagger\hat{k}}$ can be simply generated by evolving the system under the Hamiltonian $\hat{H}_p^\prime=\omega_p\hat{k}^\dagger\hat{k}$ for duration $t=\pi/\omega_p$. Physically,  this Hamiltonian can be realized by either adiabatically adjusting the trapped frequency via slowly changing the trapped voltage $V_0$(as the trapped frequency $\omega\propto V_0$~\cite{Wineland1998}), or using the parametric driving~\cite{Burd2019}  $ \hat{H}_d = g\sin(\omega t)\left(\hat{k}^\dagger+\hat{k}\right)^2$, which becomes $\hat{H}_d^\prime=g\hat{k}^\dagger\hat{k}$  under the usual RWA.
Interestingly, one can check that the spin-motion disentangled  evolution $\hat{U}_-(t)$ can be simply implemented as   $\hat{U}_-(t) = e^{-i\pi\hat{a}^\dagger\hat{a}}\hat{U}_+(t)e^{i\pi\hat{a}^\dagger\hat{a}}$.
Therefore, the PMDD technique, i.e., the evolution shown in Eq.~(\ref{psi2}), can be realized as
\begin{equation}
|\Psi_2(4t)\rangle=e^{-i\pi\hat{a}^\dagger\hat{a}}\hat{U}_+(t)e^{i\pi\hat{a}^\dagger\hat{a}}\hat{U}_+(t)e^{i\pi\hat{b}^\dagger\hat{b}}\hat{U}_+(t)e^{i\pi\hat{a}^\dagger\hat{a}}\hat{U}_+(t)|\psi_0\rangle,
\end{equation}
by setting the evolution duration as  $t=\tau$.
The relevant operational series are schematically shown in Fig.~\ref{f1}(b), with the pulse sequences; $\tau, \pi_a, \tau, \pi_b, \tau, \pi_a$, and $\tau$. Here, $\tau$ is the duration of the applied $\hat{U}_+$-operation, and $\pi_k$ is the pulse used to generate the $\pi$-phase shift operation on the $k=a,b$ mode vibration.
Therefore,  the proposed 1+2 dimensional dynamical decoupling operations are feasible with the current techniques.

For the typical parameters ~\cite{Campbell2017,Gilmore2021}, i.e., $x_d=100\mu$m (corresponds to the average phonon number $\langle\bar{N}\rangle\approx10^6$), $t=100ms$,
$\alpha=2\beta=10kHz$, Fig.~3 illustrates how the measured spin population evolves over time, with and without applying the PMDD technique.
It is seen that, without the proposed PMDD, the spin population decays rapidly, with a decay rate $R\approx\exp(-8\bar{N}_c\sin^2(\Omega T_c/2))$~\cite{Campbell2017,Wang2023}. This indicates that the spin coherent time $T_c$ decreases with the increasing rotation parameter $\Omega$ and the phonon number $\bar{N}_c$, i.e., $T_c\ll 1/(\Omega\sqrt{\bar{N}_c})$ with $\sqrt{\bar{N}_c}$ being the amplitude of cat-state.
Therefore, the effective phase accumulated time or the allowed dynamic range of the sensitive rotation sensing is severely limited~\cite{Campbell2017}, i.e., $\Omega T_c\ll1$ and thus $8\bar{N}_c\sin^2(\Omega T_c/2)\ll1$.
In contrast, with the PMDD technique proposed above, the decoherence caused by spin-motion entanglement  can be completely mitigated. As a consequence, the high  sensitivity of rotation sensing, with $\Delta\Omega\sim 3\times10^{-6}(rad/s)/\sqrt{Hz}$ for the aforementioned parameters and a one-second phase accumulating time, is achievable across a sufficiently large dynamic range: $\Omega\leq 10 rad $.
\begin{figure}\label{fp}
  \centering
  \includegraphics[width=0.38\textwidth]{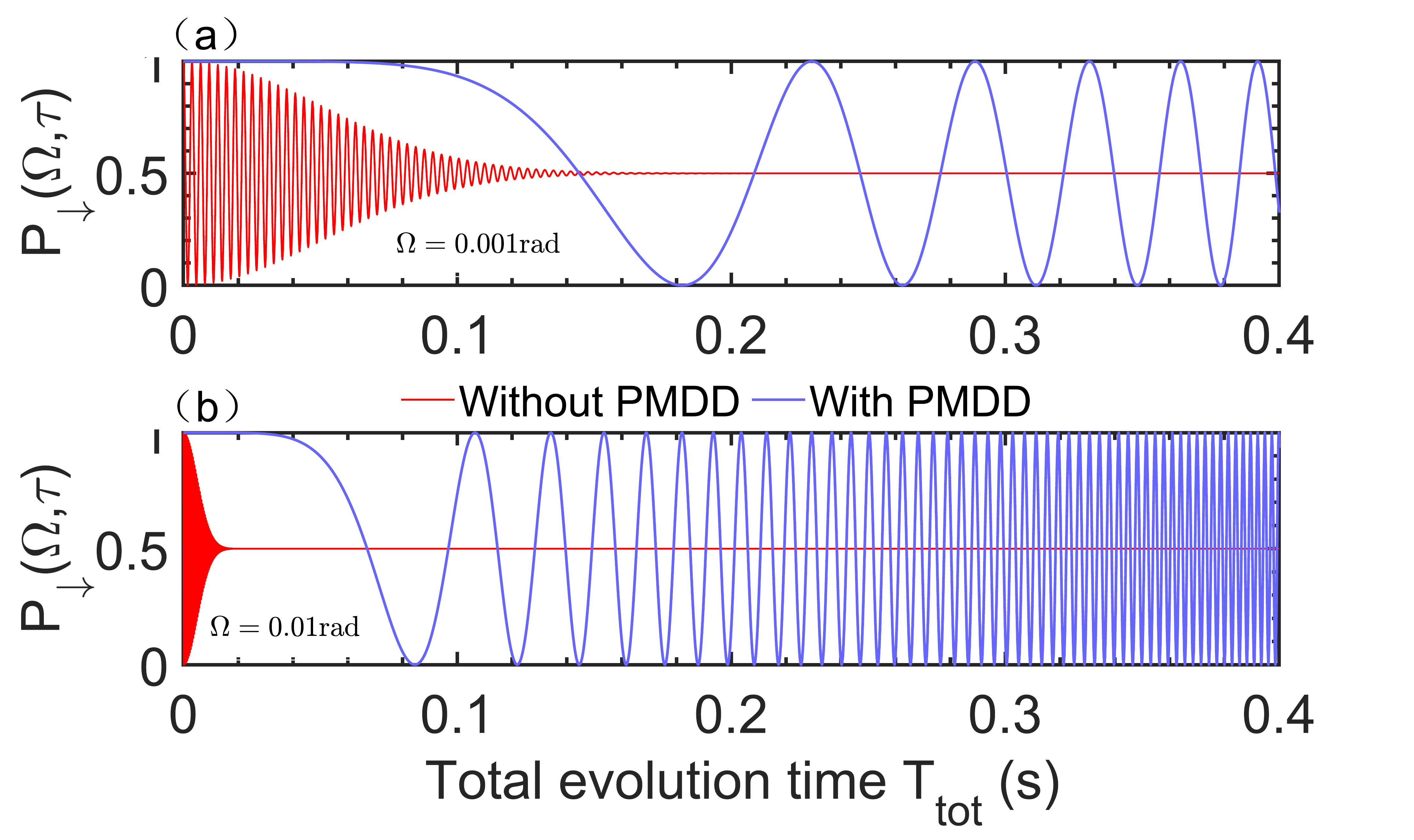}
  \caption{The spin population versus the evolution time for different rotation parameters; $\Omega=0.001$ (a) and $\Omega=0.01$ (b). Here, the blue dot-line and red-solid one refer to cases with and without the proposed PMDD technique, respectively. The comparison is made under the same condition, specifically  with the identical  entangled cat-states with $\alpha\sin(\theta)/\theta\approx 1000$.}
\end{figure}  %
Furthermore, with multiple PMDD sequences (e.g.,$M$ repetitions) the phase accumulation time can be extended further, leading to an improvement in achievable sensitivity. In fact, coherence times exceeding one hour for trapped ions have been experimentally demonstrated using dynamic decoupling (DD) techniques~\cite{Wang2021}. By fully exploiting such extended coherence times, the achievable sensitivity of rotation sensing can be accordingly enhanced $\sqrt{M}$ times. Such as, for the phase accumulating time $T_c=1000 s$ with $M\approx2500$, a sensitivity of $\Delta\Omega\sim10^{-8} (rad/s)/\sqrt{Hz}$ and a minimum detectable rotation rate $\Omega_{min}\sim10^{-10} rad/s$ could be possible with an ensemble of about 100 trapped ions~\cite{Gilmore2021}.
Additionally, unlike the case without PMDD, the dynamic range for the estimated rotation parameter $\Omega$ remains nearly unaffected~\cite{Campbell2017}.

\begin{figure}
\centering
\includegraphics[width=0.50\textwidth]{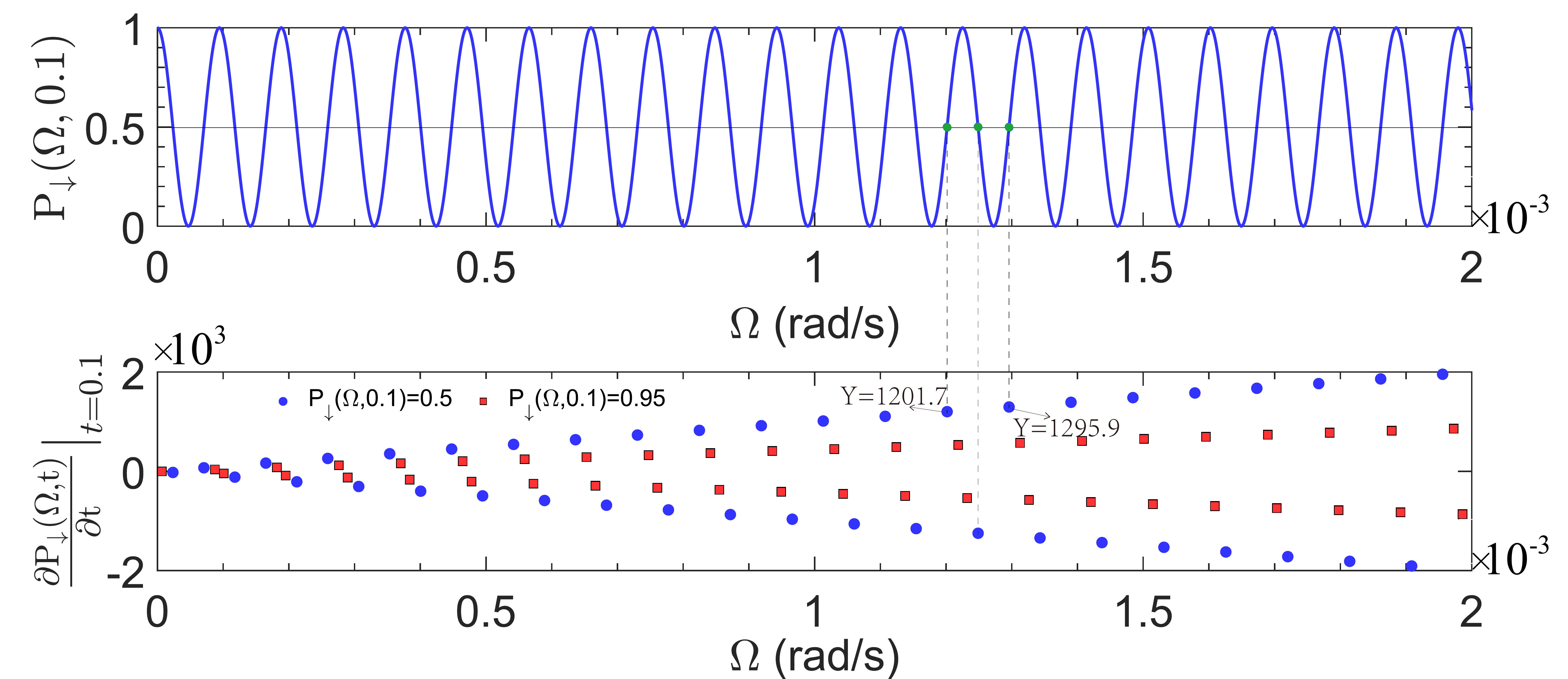}
\caption{The spin population (a) and its slopes with respective to evolution time $t$ (b), versus the $\Omega$ to show how to distinguish different  $\Omega$-parameters. Specifically, One seen that, though a population (e.g., $P_\downarrow(\Omega,0.1) = 0.5$) corresponds to numerous values of $\Omega$, the slopes (e.g., the blue dots in (b) corresponding to $P_\downarrow(\Omega,0.1) = 0.5$) could be different. Therefore, different $\Omega$-values could be distinguished.}\label{dp}
\end{figure}

Finally, a common problem of
how to determine the value of $\Omega$-parameter  unambiguously from the measurement (due to the sinusoidal variation of the spin population with respect to $\Omega$, which leads to a $2\pi$ phase ambiguity in the measurement signal (see Fig.~\ref{dp}(a)),  existing in almost all the Sagnac interferometers~\cite{Stevenson2015} and also the other quantum sensing protocols~\cite{Arai2018, Higgins2009,Liu2023}, could be resolved effectively. Indeed, For the present rotation sensing, although a specific spin population really corresponds to multiple distinct values of $\Omega$-parameter, the slopes of the measured populations, shown in Fig.~\ref{dp}, with respect to the evolution time $\tau$, however, are manifestly different, typically such as the nearest two of them could  be distinguished by approximately 100, which would be further increased with larger values of the parameter $\alpha$ and $\beta$.
Thus, different values of $\Omega$-parameter (corresponding to the same measurement $P_\downarrow(\Omega,t)$) could be unambiguously determined by comparing the values of $\partial P_{\downarrow}(\Omega,t)/\partial t$. Furthermore, the values of $\partial P_{\downarrow}(\Omega,t)/\partial t$ can be easily obtained by precisely adjusting the evolution time $t$, which is entirely feasible in experiments, in principle.

{\it Conclusions and Discussions.---}
In summary, we proposed an effective approach, called the PMDD technique, to  completely decouple the spin and the two-dimensional oscillator in a 1+2 dimensional TPSI configuration. Such a PMDD technique by using a sequence of $\pi$-phase instead of the $\pi$ pulse, is different from the conventional dynamical decoupling, which, however, can not work in the present rotation sensing. With the PMDD, the decoherence caused by the spin-motion entanglement and the other noises might be suppressed, thus allowing higher sensitivity of rotation sensing. Importantly, this high sensitivity can be achieved in the wide dynamic range, different from the previous 1+2 dimensional TPSI configuration without the complete decoupling of the spin and motion.

Certainly, although the feasibility of the proposed PMDD technique has been demonstrated using an experimental single-ion trap system, its robustness against various inevitable noise sources~\cite{Campbell2017,Wang2023}-such as decoherence caused by heating, applied field fluctuations, and anharmonicity-requires further investigation, which have been also discussed in . Additionally, it is suggested that, since the PMDD technique shares similarities with conventional dynamical decoupling (DD), it could also be used to suppress the external noise beyond that induced by spin-motion entanglement, thereby extending the coherence time of the spin. In fact, the PMDD technique can be also combined with additional $\pi$-pulses by periodically inserting them at the beginning of each evolution $\hat{U}_+(t)$, without affecting the overall results. Thus, it can be deduced that the PMDD technique also holds promise for further extending coherence times of the system, thereby achieving higher sensitivity of rotation sensing.

\section*{Acknowledgments}
This work was partially supported by the National Key Research and Development Programme of China (NKRDC) (Grant No. 2021YFA0718803) and the National Natural Science Foundation of China (NSFC) (Grant No. 11974290).

\newpage
\onecolumngrid
\appendix
\section{Derivations of the evolution operators shown in Eqs.~(3) and (5)}
In this appendix, we provide the  derivation of the evolution operator for the Hamiltonian (2), in detail. Since Hamiltonian (\ref{ah1}) is time-independent, one can assume that its time-evolution operator can be expressed as
\begin{equation}
  \hat{U}(t)=\exp\left[i\Omega tJ_y+i\alpha t\hat{x}_a\sigma_z+i\beta t\hat{p}_b\right],
\end{equation}
Using the Trotter formula~\cite{Schmidt1995}, we have
\begin{equation}\label{aevo1}
\begin{split}
\hat{U}(t)=&\lim_{n\rightarrow\infty}\left[\exp\left(\frac{i\Omega t}{n}J_y+\frac{i\alpha t}{n}\hat{x}_a\sigma_z+\frac{i\beta t}{n}\hat{p}_b\right)\right]^n\\
=& \lim_{n\rightarrow\infty}\left[\exp\left(\frac{i\Omega t}{n}J_y\right)\exp\left(\frac{i\alpha t}{n}\hat{x}_a\sigma_z+\frac{i\beta t}{n}\hat{p}_b\right)\right]^n
\end{split}
\end{equation}
It is easily to check that
\begin{equation}\label{aformula1}
\begin{split}
 &e^{i\theta J_y}\hat{x}_ae^{-i\theta J_y}=\cos(\theta)\hat{x}_a+\sin(\theta)\hat{x}_b,\hspace{3mm} e^{i\theta J_y}\hat{p}_ae^{-i\theta J_y}=\cos(\theta)\hat{p}_a+\sin(\theta)\hat{p}_b\\
 &e^{i\theta J_y}\hat{x}_be^{-i\theta J_y}=\cos(\theta)\hat{x}_b-\sin(\theta)\hat{x}_a,\hspace{3mm} e^{i\theta J_y}\hat{b}_be^{-i\theta J_y}=\cos(\theta)\hat{p}_b-\sin(\theta)\hat{p}_a,
\end{split}
\end{equation}
and thus one can obtain
\begin{equation}\label{aevo2}
\begin{split}
&\left[\exp\left(\frac{\Omega t}{n}J_y\right)\exp\left(\frac{\alpha t}{n}\hat{x}_a\sigma_z+\frac{\beta t}{n}\hat{p}_b\right)\right]^n\\
=&\Pi^{n}_{k=1}\exp\left(\frac{\alpha t}{n}\left[\hat{x}_a\cos(\frac{k\theta}{n})+\hat{x}_b\sin(\frac{k\theta}{n})\right]\sigma_z+\frac{\beta t}{n}\left[\hat{p}_b\cos(\frac{k\theta}{n})-\hat{p}_a\sin(\frac{k\theta}{n})\right]\right)\hat{P}(\theta)\\
=&e^{i\phi_1}\exp\left(\frac{\alpha t}{n}\left[\hat{x}_a\sum_{k=1}^n\cos(\frac{k\theta}{n})+\hat{x}_b\sum_{k=1}^n\sin(\frac{k\theta}{n})\right]\sigma_z+\frac{\beta t}{n}\left[\hat{p}_b\sum_{k=1}^n\cos(\frac{k\theta}{n})-\hat{p}_a\sum_{k=1}^n\sin(\frac{k\theta}{n})\right]\right)\hat{P}(\theta)\\
=&e^{i\phi_1}\exp\left(\frac{\alpha t}{n}\left[\hat{x}_a\sum_{k=1}^n\cos(\frac{k\theta}{n})+\hat{x}_b\sum_{k=1}^n\sin(\frac{k\theta}{n})\right]\sigma_z+\frac{\beta t}{n}\left[\hat{p}_b\sum_{k=1}^n\cos(\frac{k\theta}{n})-\hat{p}_a\sum_{k=1}^n\sin(\frac{k\theta}{n})\right]\right)\hat{P}(\theta)\\
\approx& e^{i\phi_1}\exp\left({\alpha t}\left[\hat{x}_a\frac{\sin(\theta)}{\theta}+\hat{x}_b\frac{1-\cos(\theta)}{\theta}\right]\sigma_z+{\beta t}\left[\hat{p}_b\frac{\sin(\theta)}{\theta}-\hat{p}_a\frac{1-\cos(\theta)}{\theta}\right]\right)\hat{P}(\theta)\\
=&e^{i(\Phi+\phi_1)}\exp\left({\alpha t}\left[\hat{x}_a\frac{\sin(\theta)}{\theta}+\hat{x}_b\frac{1-\cos(\theta)}{\theta}\right]\sigma_z\right)\exp\left({\beta t}\left[\hat{p}_b\frac{\sin(\theta)}{\theta}-\hat{p}_a\frac{1-\cos(\theta)}{\theta}\right]\right)\hat{P}(\theta),
\end{split}
\end{equation}
where
\begin{equation}
  \begin{split}
 \phi_1 &= -\frac{i}{2}\sum_{k=2}^n \left[\frac{\alpha t}{n}\left(\hat{x}_a\cos(\frac{k\theta}{n})+\hat{x}_b\sin(\frac{k\theta}{n})\right)\sigma_z+\frac{\beta t}{n}\left(\hat{p}_b\cos(\frac{k\theta}{n})-\hat{p}_a\sin(\frac{k\theta}{n})\right)\right.,\\
 &\left.\frac{\alpha t}{n}\left(\hat{x}_a\sum_{j=1}^{k-1}\cos(\frac{j\theta}{n})+\hat{x}_b\sum_{j=1}^{k-1}\sin(\frac{j\theta}{n})\right)\sigma_z+\frac{\beta t}{n}\left(\hat{p}_b\sum_{j=1}^{k-1}\cos(\frac{j\theta}{n})-\hat{p}_a\sum_{j=1}^k\sin(\frac{j\theta}{n})\right)\right]\\
 &= \sum_{k=2}^n2\frac{\alpha t}{n}\frac{\beta t}{n}\left[-\cos(\frac{k\theta}{n})\sum_{j=1}^k\sin(\frac{j\theta}{n})+\sin(\frac{k\theta}{n})\sum_{j=1}^{k-1}\cos(\frac{j\theta}{n})\right]\sigma_z\\
 & =-2\alpha t\beta t\sigma_z\frac{\sin(\theta)-\theta}{\theta^2},
   \end{split}
\end{equation}
is the phase, and
\begin{equation}
  \begin{split}
 \Phi &= \frac{i}{2}\left[{\alpha t}\left(\hat{x}_a\frac{\sin(\theta)}{\theta}+\hat{x}_b\frac{1-\cos(\theta)}{\theta}\right)\sigma_z,
 {\beta t}\left(\hat{p}_b\frac{\sin(\theta)}{\theta}-\hat{p}_a\frac{1-\cos(\theta)}{\theta}\right)\right]=0.
   \end{split}
\end{equation}
Thus, the  evolution operator shown in Eq.~(3) can be obtained.

%
Note that the evolution operator $\hat{U}_-(t)$ for the Hamiltonian (6)  can be easily obtained from $\hat{U}_+(t)$, by replacing the $\theta$ and $\alpha$ with $-\theta$ and $-\alpha$ respectively.  Furthermore,  by using the following relationship
\begin{equation}
\begin{split}
&\exp\left({\alpha t}\left[\hat{x}_a\frac{\sin(\theta)}{\theta}+\hat{x}_b\frac{1-\cos(\theta)}{\theta}\right]\sigma_z\right)\exp\left({\beta t}\left[\hat{p}_b\frac{\sin(\theta)}{\theta}-\hat{p}_a\frac{1-\cos(\theta)}{\theta}\right]\right)\hat{P}(\theta)\\
&=\hat{P}(\theta)\exp\left({\alpha t}\left[\hat{x}_a\frac{\sin(\theta)}{\theta}-\hat{x}_b\frac{1-\cos(\theta)}{\theta}\right]\sigma_z\right)\exp\left({\beta t}\left[\hat{p}_b\frac{\sin(\theta)}{\theta}+\hat{p}_a\frac{1-\cos(\theta)}{\theta}\right]\right),
\end{split}
\end{equation}
the evolution operator $\hat{U}_-(t)$ shown in Eq. (5) can be rewritten as
\begin{equation}\label{aevo2}
\begin{split}
\hat{U}_-(t)
=\hat{P}(-\theta)\exp\left({-\alpha t}\left[A_\theta\hat{x}_a+B_\theta\hat{x}_b\right]\sigma_z\right)
\exp\left({\beta t}\left[A_\theta\hat{p}_b+B_\theta\hat{p}_a\right]\right)e^{i\phi_1},
\end{split}
\end{equation}
with $A_\theta=\sin(\theta)/\theta$ and $B_\theta=(1-\cos(\theta))/\theta$.

\section{Eq.~(8) derivation}
First, it can be verified that the following relationship
\begin{equation}
\begin{split}
&e^{i\pi\hat{a}^\dagger\hat{a}}\hat{P}(\theta)\exp\left({\alpha t}\left[\hat{x}_a\frac{\sin(\theta)}{\theta}-\hat{x}_b\frac{1-\cos(\theta)}{\theta}\right]\sigma_z\right)\exp\left({\beta t}\left[\hat{p}_b\frac{\sin(\theta)}{\theta}+\hat{p}_a\frac{1-\cos(\theta)}{\theta}\right]\right)\\
&=\hat{P}(-\theta)\exp\left({\alpha t}\left[-\hat{x}_a\frac{\sin(\theta)}{\theta}-\hat{x}_b\frac{1-\cos(\theta)}{\theta}\right]\sigma_z\right)\exp\left({\beta t}\left[\hat{p}_b\frac{\sin(\theta)}{\theta}-\hat{p}_a\frac{1-\cos(\theta)}{\theta}\right]\right)e^{i\pi\hat{a}^\dagger\hat{a}},
\end{split}
\end{equation}
holds.  Next, by using
\begin{equation}\label{aevo3}
\begin{split}
        \hat{U}_+(t)e^{i\pi\hat{a}^\dagger\hat{a}}\hat{U}_+(t)=&
        e^{2i\phi_1}\exp\left({\alpha t}\left[\hat{x}_a\frac{\sin(\theta)}{\theta}+\hat{x}_b\frac{1-\cos(\theta)}{\theta}\right]\sigma_z\right)\exp\left({\beta t}\left[\hat{p}_b\frac{\sin(\theta)}{\theta}-\hat{p}_a\frac{1-\cos(\theta)}{\theta}\right]\right)\hat{P}(\theta)\\
        &\times e^{i\pi\hat{a}^\dagger\hat{a}}\exp\left({\alpha t}\left[\hat{x}_a\frac{\sin(\theta)}{\theta}+\hat{x}_b\frac{1-\cos(\theta)}{\theta}\right]\sigma_z\right)\exp\left({\beta t}\left[\hat{p}_b\frac{\sin(\theta)}{\theta}-\hat{p}_a\frac{1-\cos(\theta)}{\theta}\right]\right)\hat{P}(\theta)\\
        &=e^{2i\phi_1}e^{i\pi\hat{a}^\dagger\hat{a}}\hat{P}(-\theta)\exp\left({-\alpha t}\left[\hat{x}_a\frac{\sin(\theta)}{\theta}+\hat{x}_b\frac{1-\cos(\theta)}{\theta}\right]\sigma_z\right)\exp\left({\beta t}\left[\hat{p}_b\frac{\sin(\theta)}{\theta}+\hat{p}_a\frac{1-\cos(\theta)}{\theta}\right]\right)\\
        &\times\exp\left({\alpha t}\left[\hat{x}_a\frac{\sin(\theta)}{\theta}+\hat{x}_b\frac{1-\cos(\theta)}{\theta}\right]\sigma_z\right)\exp\left({\beta t}\left[\hat{p}_b\frac{\sin(\theta)}{\theta}-\hat{p}_a\frac{1-\cos(\theta)}{\theta}\right]\right)\hat{P}(\theta)\\
        &=e^{2i\phi_1}e^{i\pi\hat{a}^\dagger\hat{a}} \hat{U}_-(t)\hat{U}_+(t).
\end{split}
\end{equation}
and Eq.~(\ref{aevo3}),  we have
\begin{equation}\label{aevo4}
\begin{split}
      \hat{U}_+(t)e^{i\pi\hat{a}^\dagger\hat{a}}\hat{U}_+(t)e^{i\pi\hat{b}^\dagger\hat{b}}\hat{U}_+(t)e^{i\pi\hat{a}^\dagger\hat{a}}\hat{U}_+(t)
    =  e^{4i\phi_1}e^{i\pi\hat{a}^\dagger\hat{a}} \hat{U}_-(t)\hat{U}_+(t)e^{i\pi\hat{N}} \hat{U}_-(t)\hat{U}_+(t)
    = e^{4i\phi_1}e^{i\pi\hat{a}^\dagger\hat{a}}.
\end{split}
\end{equation}
Finally, applying the above evolution  sequence to the initial state $|\Psi_i\rangle$,  Eq.~(\ref{psi2}) is obtained.

\end{document}